\documentclass[transmag]{IEEEtran}
\usepackage{latexsym}
\usepackage{graphicx}
\usepackage{amsbsy,amsmath,amssymb,epsfig,bbm,mathrsfs,multirow,amsthm}

\usepackage{float}
\usepackage{cite}
\usepackage{array,multirow,graphicx}
\usepackage{graphics}
\usepackage{subfigure}
\usepackage{comment}
\usepackage{algpseudocode}

\usepackage{hyperref} 

\usepackage{color}
\usepackage{lipsum}
\usepackage[T1]{fontenc}

\def\BibTeX{{\rm B\kern-.05em{\sc i\kern-.025em b}\kern-.08em T\kern-.1667em\lower.7ex\hbox{E}\kern-.125emX}}
\usepackage{multirow} 
\usepackage{url} 

\usepackage{setspace} 

\newcommand{\modified}[1]{\textcolor{black}{#1}} 

\begin{document}

\title{From Technology to Society: An Overview of Blockchain-based DAO}

\author{
      Lu~Liu,
      Sicong~Zhou,
      Huawei~Huang\IEEEauthorrefmark{1},
      Zibin~Zheng\IEEEauthorrefmark{1}
      
      \thanks{
      This work was supported in part by the Key-Area Research and Development Program of Guangdong Province (No. 2019B020214006), the National Natural Science Foundation of China (No. 61902445, No. 61722214), the Fundamental Research Funds for the Central Universities of China (No.19lgpy222), and the Guangdong Basic and Applied Basic Research Foundation (No. 2019A1515011798).
      
      Lu~Liu and Sicong~Zhou have the equal contribution.
      Authors are all with the School of Computer Science and Engineering, Sun Yat-Sen University, Guangzhou, China.
      
      \IEEEauthorrefmark{1}Corresponding author. Email: \{huanghw28, zhzibin\}@mail.sysu.edu.cn
      }
    
}
\date{April 2020}

\IEEEtitleabstractindextext{

\begin{abstract}
Decentralized Autonomous Organization (DAO) is believed to play a significant role in our future society governed in a decentralized way. In this article, we first explain the definitions and preliminaries of DAO. Then, we conduct a literature review of the existing studies of DAO published in the recent few years. Through the literature review, we find out that a comprehensive survey towards the state-of-the-art studies of DAO is still missing. To fill this gap, we perform such an overview by identifying and classifying the most valuable proposals and perspectives closely related to the combination of DAO and blockchain technologies. We anticipate that this survey can help researchers, engineers, and educators acknowledge the cutting-edge development of blockchain-related DAO technologies.
\end{abstract}

}

\maketitle

\section{Introduction}\label{sec:introduction}

\begin{figure}[t]
\centering
\includegraphics[width=0.8\linewidth]{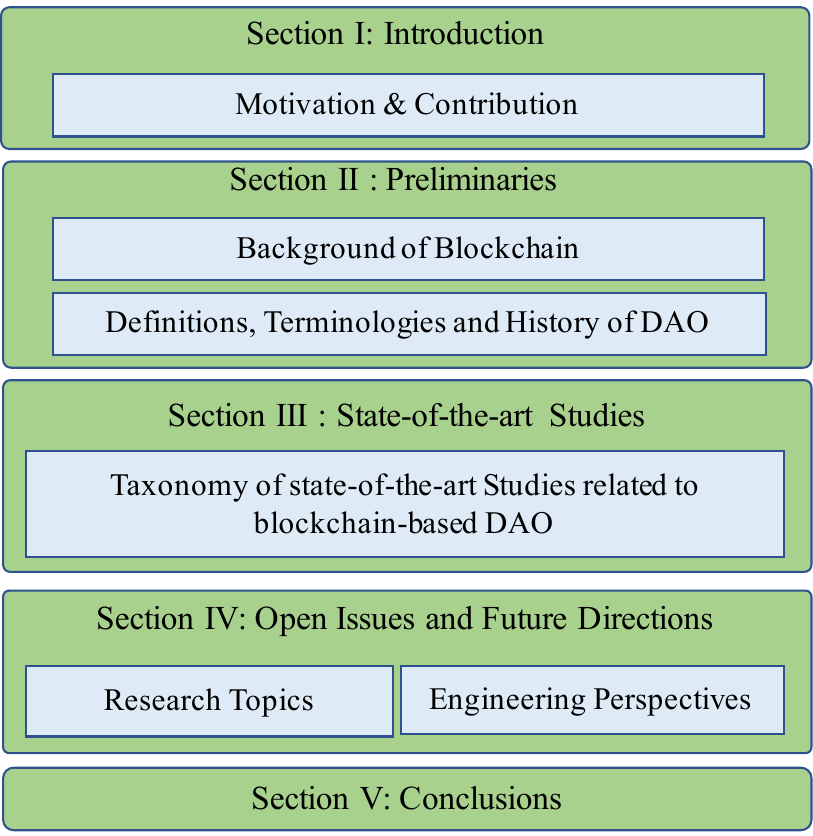}
\caption{The structure of this overview.}
\label{fig:structure}
\end{figure}

 
 {\color{black}
 
 Blockchain-based technologies have been deeply adopted by multiple applications that are closely related to every corner of our daily life, such as cryptocurrencies, tokenomics, business applications, Internet-of-Things (IoT) applications, and etc..
Decentralized Autonomous Organization (DAO), as one of the blockchain applications shown in Fig. \ref{fig:DAOnet}, is growing rapidly and drawing great attention from both academia and the governments around the world.
 Although DAO has brought a lot of opportunities to blockchain technologies, we are surprised to find out that an overview of DAO in the perspective of blockchains is still missing.
 Based on the background mentioned above, we perform a comprehensive classification on the latest studies combining the blockchain and DAO.

 The taxonomy in this article mainly includes three categories.
 In the first category, we discuss the common problems and the related studies of blockchain and DAO, including various attacks and security issues of blockchains, and the counter-trend issues. 
 In the second category, we focus on the issues related to DAO governance and the existing development of a more in-depth discussion. 
 In the third category, we evaluate the latest development of DAO in various fields, such as e-government, economy, and etc., and predict the future development directions of the relevant fields.

\begin{figure}[t]
\centering
\includegraphics[width=0.68\linewidth]{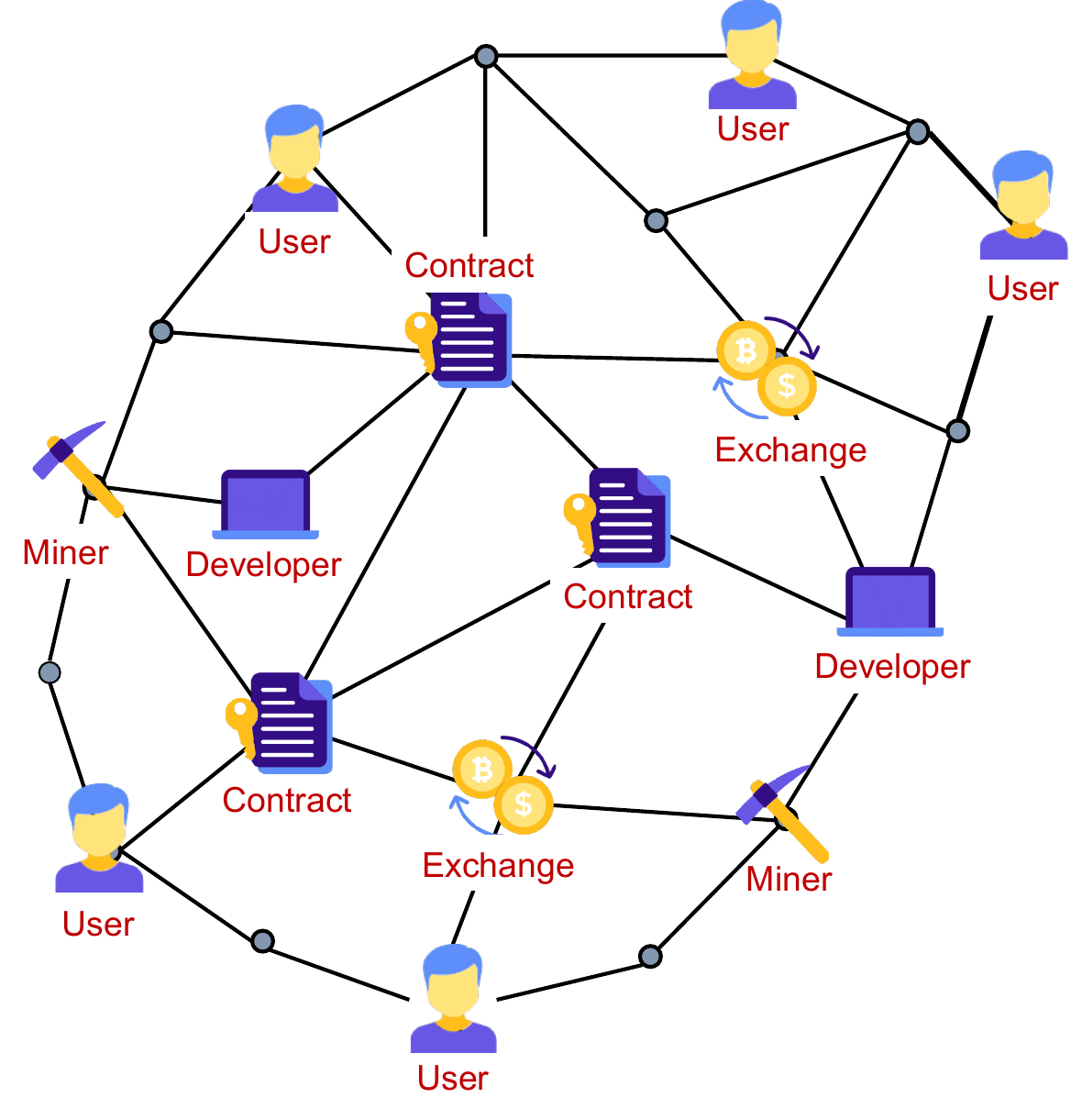}
\caption{The structure of DAO.}
\label{fig:DAOnet}
\end{figure}

 Although a small number of researchers are worried about the future of DAO and blockchain due to the \textit{hard fork} caused by \textit{The DAO} hacking incident \cite{dhillon2017dao}, most of them still have high expectations on this technology.
 We also believe that with the inspiring pace of the development of DAO and blockchain, the DAO projects can become mature, and DAO will show overwhelming advantages over many existing solutions.

 Through this overview, we find the most important findings on this direction is that the future effort can be devoted to improving DAO from a better balance of decentralization, security and scalability. We look forward to a new integration of both social and organizational structure of DAO in the context of blockchain technologies.
 To help have a clear clue of this article, Fig. \ref{fig:structure} shows the organization of this article.
 
 }

\section{Preliminaries}


 \subsection{Decentralization}

 {\color{black}
 
 In general case of centralization, the use of a database is basically based on the trust of a third-party organization. For example, as the third party, people all trust the banking system, which can  correctly manage the database and record our transactions. The bank keeps accounts for every transaction, and only the bank has the authority to keep users' accounts. However, the shortcoming of such centralized organization is obvious. No one can make sure that whether the centralized organization that manages the database is entirely trustworthy. For example, during the global economic crisis in 2008, the US government could issue money indefinitely, because it is the central institution of monetary management.

On the contrary, decentralization means that the database does not depend on a specific organization or administrator but is distributed among all peers. 
Blockchain is essentially a decentralized database.
Each full node has a complete copy of the blockchain ledger. If the database is modified, the information saved by all nodes will be noticed. Thus, the information in the blockchain database will be open and transparent. Decentralization solves the trust problem through redundant data validation.

}

 \subsection{Bitcoin Blockchain}
 
 {\color{black}
 
    Originally proposed by Satoshi Nakamoto \cite{casino2019systematic}, Bitcoin is the first application of the blockchain. In Bitcoin, the blockchain serves as a distributed database that stores all transactions originated from one account to another. The advantages of blockchain brings many pros for the Bitcoin ecosystem. For example, everyone has the right to validate accounts, the currency cannot be overissued, and the entire ledger is completely open and transparent.
    When processing a transaction, Bitcoin adopts the techniques of digital signatures to identify the ownership of a coin. Each Bitcoin account address has a private and a public keys. The private key is private and is used to exercise the bitcoin ownership of the bitcoin account, while the public key is known to all nodes to verify the transaction and the balance of a Bitcoin account. When the transfer information of a transaction is published, a digital signature must be embedded by encrypting the digital summary of the transfer message together with the private key of the sender. Thus, other nodes can use the public key of the sender's account to decrypt and verify the legality of the transaction. After such verification, each blockchain node is acknowledged this transaction.
    
    }

 \begin{figure}[t]
\centering
\includegraphics[width=0.9\linewidth]{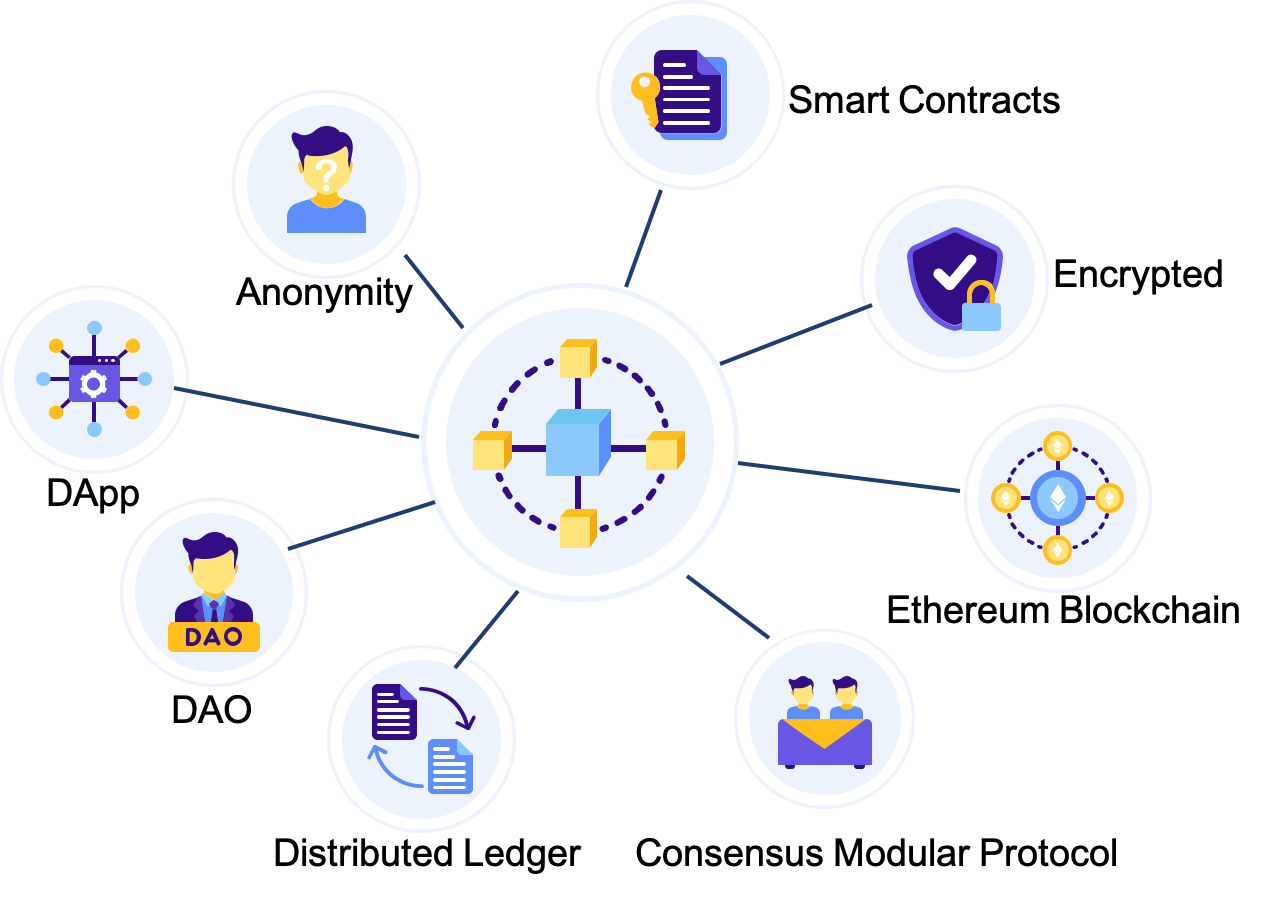}
\caption{Blockchain-related fields.}
\label{fig:blockchain}
\end{figure}

 \subsection{Definition and Background of DAO}
 
 {\color{black}
 
 DAO was originally introduced by the white paper \cite{buterin2014ethereum}, in which DAO is defined as an organization built on smart contracts that can execute autonomously. Unlike the conventional centralized entities, it doesn't include the central control or management.
 As shown in Fig. \ref{fig:DAOnet}, DAO achieves the decentralized organization by encoding a set of rules in smart contracts, where how DAO performs is predefined. Although this completely decentralized way makes investors typically don't know and trust each other, blockchain is a good tool to help achieve the goal of DAO.
 However, DAOs are not necessarily to built on top of an existing blockchain such as Ethereum \cite{wood2014ethereum}.
During the getting-start guidelines, Casino \cite{casino2019systematic} was published as an introduction to DAO. It presents the concepts, characteristics, frameworks, applications, the future trends of DAO, and etc. The readers will gain a systematic overview of DAO that spans multiple domains.
 }

 \subsection{Project The DAO}
 
 {\color{black}
 
In the development of DAO, the first DAO project is proposed with a historic and dramatic meaning. A particular historical moment of DAO is the creation of the first DAO and how it was eventually hacked. 
\textit{The DAO} project was started on April 30th, 2016. By the end of the entire funding period, more than 11,000 enthusiastic members had participated and raised 150 million dollars, making the \textit{The DAO} the largest crowdfunding project in history. It was an overnight success, but the idea of \textit{The DAO} vulnerability had been circulating in the developer community.
Finally on June 18th, a hacker began using a ``recursive call vulnerability" in the software to collect Ether coins from The DAO's token sales. He took advantages of a well-intentioned but poorly implemented feature of DAO designed to prevent the majority from tyrannizing over dissenting DAO token holders. But this ``split" feature was implemented to make \textit{The DAO} vulnerable to catastrophic reentrance errors. As a result, the attacker was able to steal about 3.6 million ETH, worth about \$50 million at the time of the attack, bringing the price of the coin from more than 20 dollars to less than 13 dollars. Losses reached 70 million dollars. \textit{The DAO}'s problems have had a negative impact on both the Ethereum network and its cryptocurrency. 
The situation for \textit{The DAO} investors is particularly precarious. Eventually, over 90\% of the hashrates indicated support for a fork. \textit{The DAO} fund was returned to investors as if the organization never existed. \textit{The DAO} hack sparked a debate about hard forks and soft forks, and the creation of Ethereum Classic \cite{dhillon2017dao}.
}

 \subsection{To Launch a DAO Project}

\begin{figure}[t]
\centering
\includegraphics[width=0.9\linewidth]{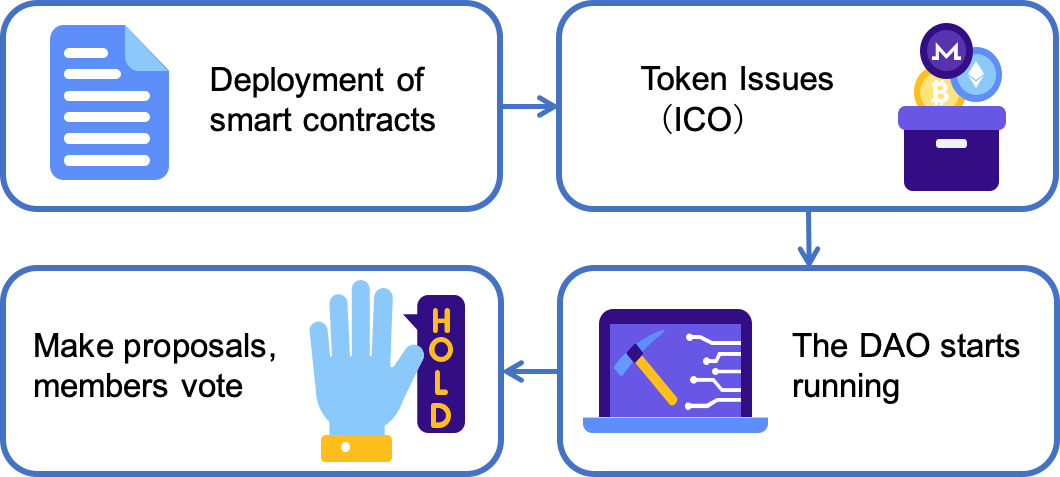}
\caption{The four-step launch of a DAO project.}
\label{fig:a_DAO_project}
\end{figure}

 {\color{black}

Although the first DAO project failed, it did not completely prevent the initiation and development of other DAO projects. The operation of launching a DAO project includes the following steps \cite{ProjectSteps}, which are also illustrated in Fig. \ref{fig:a_DAO_project}:
\begin{enumerate}
\item Developing and deploying smart contracts according to predefined rules.
\item Handling the token issues (through ICO) at the initial financing stage.
\item At the end of the financing phase, a DAO starts running.
\item Proposals are made and members can vote on them. 
\end{enumerate}

 }

 \subsection{Existing Popular DAOs}
 
 The representative existing implementations of DAO are reviewed as follows.
 
 \subsubsection{Aragon}
 
 {\color{black} 
 
 Aragon \cite{aragon2020} is a platform that enables any participant to collaborate with others without any third-party organizers. On Aragon, people can create a decentralized digital jurisdiction for their company, community, and organization. The Aragon users are also allowed to create diverse communities. For example, a financial DAO can be generated to incentivize internal usage of coninbase Apps, a council DAO is launched to expand the utilization of wealth management, or a pocket DAO is established to eliminate blockchain infrastructure monopolies.

 \subsubsection{Colony}
    
  Colony \cite{colony2020} is designed as an infrastructure that enables organizations collaborate with each other via the decentralized software implemented on top of Ethereum. It treats every participant impartially. Unlike other DAOs, Colony advocates to eliminate the requirement of voting from members. Instead, it focus on mechanisms that enforce people to get their job done.
    
 \subsubsection{DAOstack} 
 
 DAOstack \cite{DAOstack} is an open-source, modular DAO project, which leverages the technology and adoption of decentralized governance, enabling people to create the DApps (decentralized apps), DAOs, and DAO tools.

 
  Although several DAOs have been implemented, we should notice that DAO is still in its immature stage, a large number of new DAOs will be developed for future, and many insights of DAOs will be perceived then.
 }

\section{Taxonomy of State-of-the-art Studies}\label{sec:taxonomy}
 
 {\color{black} 
 
 In this part, we perform a thorough taxonomy towards the existing up-to-date studies of blockchain-related DAO. Through the classification of the existing literature related to DAO, we find that in addition to the introduction of DAO systems, existing studies mainly focus on the security issues, applications in various fields, and DAO governance issues. Therefore, we divide it into the categories presented in each subsection.

 Among those categories, on the analysis of the questions about DAOs related issues, we divide them into two classes. On one hand, a DAO project is based on the blockchain, thus the problems existing in blockchain are also the existing problems of DAO. On the other hand, DAO induces some new problems in governance. Such relevant studies are reviewed in governance problems and solutions.
 Finally, in the last category, DAO and related areas, we pay attention to the current development of DAO in the context of various fields.

 }


\subsection{Existing Problems and Solutions of Blockchains}

{\color{black}

 The hacking incident of the first \textit{The DAO} project triggered the introspection of DAO project \cite{santos2018dao}. 
The first generation blockchain technology, Blockchain 1.0, was mainly invented for cryptocurrency purposes. Then, the second generation of Blockchain 2.0, represented by Ethereum, is an open platform that enables a new decentralized computing paradigm. The DAO is exactly based on Ethereum. While there are no obvious security vulnerabilities in pure cryptocurrency systems such as Bitcoin \cite{chen2020survey}, the second-generation blockchain applications and semantics inevitably introduce security vulnerabilities \cite{7906988, 8369416}.

}

%
{\color{black}
Besides, from the perspective of social development, we observe some common problems that need to be solved together for both blockchains and the DAO. For example, the typical problems include:
\begin{itemize}
    \item the interpretation of fork culture by DAO hacker event;
    \item whether there are problems in the application development of distributed ledger technology (DLT);
    \item and whether the main trend of blockchain technology is reasonable.
\end{itemize}
 Inspired by those questions, the blockchain system hacker attacks \cite{dhillon2017dao} and security issues \cite{li2017survey}, blockchain counter-trend issues \cite{manski2017building} have drawn a lot of attention.
}

\begin{table*}[h]
\caption{Existing DAO as blockchain problems and solutions}
\centering
\footnotesize
\begin{tabular}{|p{0.15\textwidth}|p{0.2\textwidth}|p{0.4\textwidth}|}%
\hline
\textbf{References} &\textbf{Recognition} &\textbf{Methodology}\\
\hline
	 
	Li \cite{li2017survey} & Security threats and enhancement solutions in blockchain & A systematic study on the security threats to blockchain and the corresponding real attacks, and suggests some future directions to stir research efforts into this area.\\

	\hline
	Zhou \cite{zhou2020solutions} & Scalability of blockchain & Existing blockchain scalability solutions are classified according to the blockchain hierarchy.\\
	
	\hline
	Manski \cite{manski2017building} & Countervailing trend & Blockchain applications could exacerbate inequality.\\
	
	\hline
	LSE Team \cite{LSE2018} & DLT & The potential of DLT is great but need to be assessed the feasibility.\\

	\hline

\end{tabular}
\label{Table:Existing}
\end{table*}

\subsubsection{\modified{Security threats and enhancement solutions in blockchain}}

{\color{black}

Blockchain technology has shown a promising application prospect since its birth.
Blockchain has been used in many fields, ranging from the original cryptocurrency to various applications based on smart contracts.
Along with the booming development of blockchains, the security and privacy of blockchains should not be ignored.
On the basis of existing studies on blockchain security and privacy issues, Li \textit{et al.} \cite{li2017survey} systematically studied the security threats of blockchains through the analysis of popular blockchain systems. Their major contribution includes (a) analyzing the causes and possible consequences of each risk or vulnerability, (b) investigating the corresponding actual attack, and (c) analyzing the exploited vulnerability.

From the generation perspective of blockchains, we summarize the common risks of the blockchain 1.0 as follows: (a) 51\% vulnerability, (b) Private key security, (c) Criminal activity, (d) Double spending, and (d) Transaction privacy leakage.
While for blockchain 2.0, the common risks include: (a) Criminal smart contracts, (b) Vulnerabilities in smart contracts, (c) Under-optimized smart contracts, and (d) Under-price operations.

Furthermore, the popular attacks towards the blockchain include:  Selfish mining attacks,  DAO attacks (which is also the focus of our article),  BGP hijacking attacks,  Eclipse attacks,  Liveness attacks, and Balance attacks. 
Considering those attacks, Li \textit{et al.} \cite{li2017survey} summarized the security-enhancement solutions of the blockchain system as follows: (a) SmartPool,
 (b) Quantitative framework, 
 (c) Oyente, 
 (d) Hawk, and (e) Town Crier. Those solutions have made a good prediction for the future direction of the blockchain.

}

\subsubsection{\modified{Scalability of blockchain \cite{zhou2020solutions}}} 

{\color{black}

Similar to CAP theory in the field of traditional distributed systems, the three important attributes of blockchain systems, including decentralization, security, and scalability, cannot be fulfilled altogether. For example, Bitcoin faces performance problems with low throughput and high transaction latency. Other cryptocurrencies also face these flaws, leading researchers to pay more attention on the scalability of blockchains.
To have a clear clue about the blockchain scalability solutions, Zhou \textit{et al.} \cite{zhou2020solutions} attempted to overview the related state-of-the-art studies by categorizing them according to their blockchain hierarchy. The hierarchical structure mainly consists of two layers.
The first-layer solution is executed on the chain, focusing on the blockchain consensus, networks and data structures. Such as increasing the block size of Bitcoin blockchain, optimizing the storage scheme, as well as adopting the sharding technology. Various improved consistency algorithms, where transaction throughput can be increased and transaction latency is decreased, are also reviewed. 
The second-layer solutions seek opportunities to extend the blockchain through the off-chain channels, side-chain and cross-chain protocols. Basically, these solutions have both advantages and limitations as they strive to achieve decentralization, security, and scalability at the same time. The insightful classification and analysis of current solutions can inspire further researches.

}

\subsubsection{\modified{Countervailing trend}}

{\color{black}

To our surprise, the blockchain technology also somewhat shows a counter-trend problem.
Blockchains, like other technologies, have shown a tendency to pursue different future trajectories, depending on their implementation details.
For example, blockchain technology can help build a technology community in which advanced exchanges, communications, and decision-making technologies are used to aggregate, allocate, and manage the capital at multiple levels. 
However, a series of anti-subsidy trends indicate a deepening of inequality and democratic decline. Because technology is stratified, a large number of employees are reduced to less disposable population, regulation is reduced and corporate personnel are technologized. 
While the mainstream trends in blockchain technology are greatly believed as distribution, decentralization and democratization, the most powerful blockchain applications are likely to exacerbate inequality.

}

\subsubsection{\modified{The future of the DLT}}

{\color{black}

In fact, blockchain is essentially an application of the distributed ledger technology (DLT). LSE Team \cite{LSE2018} describes the problem to be solved before DLT is used.
The potential of DLT is too great to ignore. Its commitment to decentralization, data security and privacy can help improve and make public services more affordable by reducing the role of government as an intermediary.
The decentralized and transparent properties of the DLT lead to greater collaboration and integration with the private and social sectors by enhancing the government's own transparency, accountability and inclusiveness.
However, there is no general rule book that specifies where a DLT should be deployed. If governments desire to use DLT for governance, they need to assess the feasibility of DLT and implement the DLT-based applications only when the benefits of speed, security and privacy outweigh the social costs. 
Governments must govern in the context when the implemented DLT applications are transparent to the underlying algorithms and ensure that the applications truly represent public value. All these visions will have to wait for the DLT technologies to evolve further.

}


\subsection{Governance Problems and Solutions}

{\color{black}

 DAO, as an application of blockchain in governance, raises new issues about governance that are not found in conventional blockchain. 
 Through professional knowledge and experiments \cite{tarasiewiczforking}, there is still existing differences between code-based governance and blockchain-based governance. To use DAOs in certain areas, DuPont\cite{dupont2017experiments} argues that DAO may simply be a risky investment which masquerads as a new way of doing things.
 Furthermore, once governance is applied in the area of blockchains, legal problems are inevitable. In particular, the characteristics of blockchain governance have led to tensions between strict ``on-chain" governance systems and possible ``off-chain" governance applications \cite{reijers2018now}. Before \textit{The DAO} attack, some lawyers expressed concerns about DAO programs, saying that DAOs have touched on legal issues related to security in several countries \cite{blemus2017law}.
 
 }

\begin{table*}[h]
\caption{Governance problems and solutions}
\centering
\footnotesize
\begin{tabular}{|p{0.1\textwidth}|p{0.1\textwidth}|p{0.13\textwidth}|p{0.44\textwidth}|}%
\hline
\textbf{Categories} &\textbf{References} &\textbf{Recognition} &\textbf{Methodology}\\
\hline
	 
	Fork & Tarasiewicz \cite{tarasiewiczforking} & fork and culture & The author interprets fork culture based on The DAO hacker event.\\
	 
	\hline
	{Governance in company}
	& Kaal \cite{kaal2019blockchain} & corporate governance & Problems with DAOs used in corporate governance.\\
	
	\cline{2-4}
	{ } & Lafarre \cite{lafarre2018blockchain} & DAO in AGM & Problems with DAOs used in corporate governance(especially AGM).\\
	
	\hline
	{Economy}
	& Beck \cite{beck2018governance} & Blockchain economy & DAO could lead to blockchain economy.\\
	
	\cline{2-4}
	{ } & Massacci \cite{massacci2017seconomics} & security and economics vulnerabilities & The failure of a security property, e.g. anonymity, can destroy a DAOs because economic attacks can be tailgated to security attacks.\\
	
	\hline
	{Law}
	& Blemus \cite{blemus2017law} & blockchain laws & This paper describes blockchain regulations discussed in the US, EU and major economic countries.\\
	
	\cline{2-4}
	{ } & Reijers \cite{reijers2018now} & DAO in legal philosophy & This paper seeks to situate the blockchain discussion within the field of legal philosophy, examining how legal theory can apply in the context of blockchain governance.\\
	
	\cline{2-4}
	{ } & Shakow \cite{shakow2018tao} & DAO tax issues & This article explains how a decentralized autonomous organization operates and interacts with the U.S. tax system.\\

	\hline

\end{tabular}
\label{Table:governance}
\end{table*}

\subsubsection{\modified{Fork and culture for DAO}}

{\color{black}

Interpreting fork culture based on \textit{The DAO} hacker event, Tarasiewicz \textit{et al.} \cite{tarasiewiczforking} argue that a strong emphasis must be placed on interaction and communication between institutions and informal coding communities to further research and develop new blends of social and organizational structures. 

}

\subsubsection{\modified{Problems with DAOs used in corporate governance}}

{\color{black}

Agency theory is the dominant theory of governance conflicts among shareholders, company managers and creditors, in which one party entrusts the work to the other party. 
The core agency conflict caused by the separation of ownership and control cannot be fully resolved under the existing theoretical and legal framework. Attempts to monitor agents are inevitably costly and transaction costs are high. 

Kaal \textit{et al.} \cite{kaal2019blockchain} point out that blockchain provides an unprecedented solution to the agency problem in corporate governance. }{\color{black}DAO technology could help improve the agency relationship, but also proposed the potential of blockchain technology as an emerging technology of governance design, in which many ideal models and theoretical evaluations were limited by the real world. 
First of all, the blockchain is a fundamental technology, and its transformative impact will take decades rather than years to establish and reform the system.
In the corporate governance environment, the application of blockchain technology may develop in the existing centralized structure or decentralized environment. The former requires consensus on how and when to implement such technologies for governance use cases. For the latter, only when a true decentralized common blockchain emerges, with scalability and full security, can the proxy relationships be truly removed to overhaul them. 

With the complexity of the agency relationship, human behavior in the agency relationship needs a backstop, namely the continuous support of human code. Without decentralized human support for code, the immutability of the blockchain and its cryptography security systems may not create true transactional guarantees and trust between principals and agents to maintain the integrity of their contractual relationships. 

The blockchain-based corporate governance solution in DAO requires an incremental blockchain governance protocol. Thus, the socially optimal hard fork rule may not be applicable.

}

\subsubsection{\modified{Problems with DAOs used in AGM}} 

{\color{black}

The classic Annual general Meeting (AGM) has three functions for shareholders: information, forum and decision-making. AGM also has important theoretical significance in the collective supervision of shareholders. However, AGM is often regarded as a dull and obligatory annual ceremony, and all three functions are actually eroded. For example, (almost) all information is often disclosed well before the AGM. In addition to expensive shareholder voting decisions, the decision-making function of AGM also has static annual characteristics. Besides, the annual general meeting also has procedural defects. Especially when shareholders vote remotely, there is great uncertainty about whether the information between shareholders and the company (including shareholder voting records) is correctly communicated. 

Lafarre \textit{et al.} \cite{lafarre2018blockchain} therefore strongly call for the use of blockchain technology to modernize the AGM. Blockchain technology can significantly reduce the cost of shareholder voting and corporate organization. And it also can improve the speed of decision-making, and promote the rapid and efficient participation of shareholders.

}

{\color{black}

In calls for using blockchain technology to realize the modernization of the Annual general Meeting (AGM), however, at the same time, Lafarre \textit{et al.} \cite{lafarre2018blockchain} are also pointed out that it is important to realize the annual general meeting of shareholders based on blockchain will bring important legal issues. Whether it should be abolished physical classic annual general meeting of shareholders, or let it outside of the annual general meeting of shareholders based on blockchain coexist? If it is desirable to organize a decentralized AGM only on the blockchain, how many forum functions are included in this technology? Record dates and notice periods must be reconsidered, and the role of intermediaries in (cross-border) chains. More importantly, are shareholders and companies ready to participate in non-entity meetings? Recent evidence shows that even most institutional investors do not favor full virtualization.

}

\subsubsection{\modified{Blockchain economy's rethink}} 

{\color{black}

Following the development of blockchain, blockchain economy is on the rise, also needs new governance methods. 

}

{\color{black}

Blockchain and the smart contract supported by blockchain may give birth to a new economic system, which Beck \textit{et al.} \cite{beck2018governance} call blockchain economy. 
The blockchain economy goes beyond the digital economy because agreed transactions are executed autonomously, by following rules defined in smart contracts, without the need for agency intervention or third-party approval. They can embed digital assets or tokens into digital representations of physical assets to enforce autonomous contract performance. The blockchain ensures that contracts are honored without being broken. It is in the new form of organization as DAO that blockchain economy will manifest itself.

}

{\color{black}

Beck \textit{et al.} \cite{beck2018governance} explored a DAO case, Swarm City to explore the decision right, responsibility and incentive mechanism related to governance. 
Decision rights involve controlling certain assets, monitoring decisions, and have to do with accountability and incentives motivate agents to take action. 
The authors used a novel IT governance framework to show that the emergence of the blockchain economy requires a rethink of governance. Compared with the digital economy, the location of decision-making power in the blockchain economy will be more decentralized. The accountability system in principle will be more and more established technically rather than institutionally, and the consistency of incentives will become more and more important. 

Therefore, from the three governance dimensions, the authors proposed the governance research agenda in the blockchain economy in each dimension respectively. For example, in the aspect the decision right, the research agenda include: (a) how to make decisions in the blockchain economy? (b) how to allocate decision management right and decision control right? (c) how to resolve decision-making differences in the blockchain economy? and (d) what is the role of ownership in the blockchain economy?

The work \cite{beck2018governance} has identified an important approach to governance research in the blockchain economy and provided a rich foundation for further theoretical work. 

}

\subsubsection{\modified{Security and economics vulnerabilities}} 

{\color{black}

Traditionally, security and economics functionalities in IT Financial services and Protocols (FinTech) have been separate goals. Only security vulnerabilities in security-critical systems that could be exploited by terrorists, criminals or malicious government actors turned into security problems.

Massacci \textit{et al.} \cite{massacci2017seconomics} believe that security and economics are crucial issues for DAOs. \textit{The DAO}'s hack is essentially a combination of a security vulnerability (recursive calls constantly extract coins from \textit{The DAO}) and an economic attack (The user is only authorized to withdraw money the first time). In DAO Futures Exchange, on one hand, a failure of integrity could be dramatic for the agreement. On the other hand, if anonymity fails, futures exchange DAO may face economic attacks combining anonymity failure and price discrimination.

Failure of security attributes, such as anonymity, can destroy DAOs because economic attacks can be considered security attacks. The danger is not the vulnerabilities themselves, but the combination of an attack on software and an attack on the economy. Economic vulnerabilities, presumably, cannot be repaired, since the economic damage they may cause is unlikely to be reversed by pure technology such as forks. Thus for DAOs, economic vulnerabilities (security and economic vulnerabilities) are indeed the new ``beast" to be ignored.

}

\subsubsection{\modified{Infancy of blockchain laws}} 

{\color{black}

Blockchain has become a major topic for public policymakers around the world. As this disruptive and decentralized technology has become a key business issue for start-ups and market participants, central banks and financial regulators, particularly in the US and EU, have shifted from initially intense hostility to a more cautious and market-friendly stance. 

Blemus \textit{et al.} \cite{blemus2017law} collected and compared regulatory trends in various applications or issues within the area of United States, the European Union, and other key countries. These were supported by blockchain technologies, including Bitcoin/virtual currency/cryptocurrencies, smart contracts, decentralized autonomous organizations, initial coin offering (ICO), and others. It mainly includes three supervision projects: (a) supervision of virtual currency, (b) supervision of ICO (and cryptocurrency), and (c) legal validity of blockchain technology and intelligent contract.

The conclusion shows that distributed ledger technology regulation is in its infancy. As of 2017, when the work \cite{blemus2017law} was published, uncertainty remains about the legal and economic qualifications of virtual currencies, tokens, ICOs, smart contracts and distributed ledger technologies. Predictably, over time, the need for extensive research into blockchain technology will become less controversial.

}

\subsubsection{\modified{Discussion of DAO in legal philosophy}}

{\color{black}

Recently, the blockchain developer community has begun to turn its attention to governance issues. The governance of blockchain-based systems typically consists of various rules and procedures that can be implemented ``on-chain" and ``off-chain". On-chain governance refers to the rules and decision processes that are directly encoded into the underlying infrastructure of blockchain-based systems.  Off-chain governance includes all other rules and decision-making processes that may affect the operation and future development of blockchain-based systems. The characteristics of blockchain governance raise the issue of possible tensions between a strict ``on-chain" governance system and possible ``off-chain" governance applications. 

Through investigations, Reijers \textit{et al.} \cite{reijers2018now} believe that chain governance and \modified{Kelsen's positivist concept of legal order \cite{golding1961kelsen}} have a striking similarity. Blockchain-based systems become vulnerable when private interest groups use off-chain mechanisms to usurp governance systems on the chain. \textit{The DAO} attack shows that while ``code rules" can be formally followed in a specific chained order, in exceptional states, sovereignty is asserted through a chained mechanism. 

As reflected in \modified{Kelsen's argument}, the combination of private interests is a weakness of positivist legal systems, and it can be assumed that this is an inherent weakness of governance on the chain of existing blockchain-based systems. Given these characteristics, future research could consider possible steps the blockchain community could take to address abnormal states in a manner consistent with their respective ideologies. 

}

\subsubsection{\modified{DAO tax issues}}

{\color{black}

After \textit{The DAO} hack, the Ethereum community voted to create a ``hard fork" for the Ethereum chain, creating two Ethereum chains in the future. To add insult to injury, Securities and Exchange Commission (SEC) used this DAO to explain for the first time its view that some blockchain-related issuance would be considered securities subject to SEC regulation. 
The possibility of using smart contracts to allow entities to operate entirely autonomously on blockchain platforms seems attractive. 
It is not hard to see that these structures of DAO raise significant tax issues. However, little thought has been given to how such an entity would qualify for the tax system. Thus, Shakow \textit{et al.} \cite{shakow2018tao} explains how a decentralized autonomous organization operates and interacts with the U.S. tax system by describing how a DAO works, and raises many of the tax issues raised by these structures. 

As a result, there is no evidence that DAOs have considered being subject to various requirements under the tax code. For those who want to comply, the easy solution is to use a site like Overstock.com. If they don't, they may be penalized by the IRS. However, without international cooperation and innovation, it is difficult for tax administrators to find out who should tax a ``DAO" income. 

}


\subsection{DAO Technologies and the Related Areas}

{\color{black}

It is nearly undeniable that DAO still has a certain trend across diverse sectors such as supply chain, business, healthcare, IoT, privacy, and data management \cite{beck2018governance, zichichi2019likestarter, dai2017toward, jeong2018blockchain}. The emerging DAO is on the rise. The work of Beck \textit{et al.} \cite{beck2018governance} and other papers have discussed in the fields of blockchain economy, crowdfunding, accounting, and even electric cars and charging station billing systems. Especially, more than one studies \cite{beck2018governance} \cite{zichichi2019likestarter} have found that in addition to e-government, the blockchain in the financial industry is also very promising.

}

\begin{table*}[h]
\caption{DAO And Related Areas}
\centering
\footnotesize
\begin{tabular}{|p{0.1\textwidth}|p{0.1\textwidth}|p{0.13\textwidth}|p{0.44\textwidth}|}%
\hline
\textbf{Categories} &\textbf{References} &\textbf{Recognition} &\textbf{Methodology}\\
\hline
	 
	Governance in company & Leonhard \cite{leonhard2017corporate} & DAO corporation & A virtual DAO corporation with a corporate structure similar to the structure of a modern corporation. \\
	 
	\cline{2-4}
	{ } & Lumineau \cite{lumineau2020blockchain} & DAO corporation & DAO governance works differently than traditional contractual and relational governance.  \\
	 
	\hline
	{eGov DAO}
	& Diallo \cite{diallo2018egov} & DAO in e-government system & They provide a concrete use case to demonstrate the usage of DAO e-government and evaluate its effectiveness.\\
	
	\cline{2-4}
	{ } & Jun \cite{jun2018blockchain} & DAO replacing existing social apparatuses and bureaucracy & Blockchain creating ``absolute law"  makes it possible to implement social technology that can replace existing social apparatuses including bureaucracy.\\
	
	\hline
	{Economy}
	& Akgiray \cite{akgiray2019potential} & DAO tokens and money & The question of whether DAO tokens are money or not may can be raised.\\
	
	\cline{2-4}
	{ } & Zichichi \cite{zichichi2019likestarter} & crowdfunding and DAO, LikeStarter & LikeStarter is structured as a DAO, that fosters crowdfunding, and recognizes the active role of donors, enabling them to support artists or projects, while making profits.\\
	
	\hline
	{Accounting}
	& Dai \cite{dai2017toward} & blockchain in accounting profession & Blockchain has the potential to transform current auditing practices, resulting in a more precise and timely automatic assurance system.\\
	
	\cline{2-4}
	{ } & Karajovic \cite{karajovic2019thinking} & blockchain in accounting profession & A analysis of the implications of blockchain technology in the accounting profession and its broader industry. Criticisms will be raised to address concerns regarding blockchain's widespread use. \\
	
	\cline{2-4}
	{ } & Jeong \cite{jeong2018blockchain} & billing system & This paper proposes the blockchain based billing system. The EV and the charging station store the billing information in the blockchain  and prevent the modification.\\
	
	\hline
	{Voting}
	& Zhang \cite{zhang2018privacy} & DAO in voting & This paper proposes a local voting mechanism based on blockchain.\\

	\hline
\end{tabular}
\label{Table:DAO related areas}
\end{table*}

\subsubsection{Corporate Governance on Ethereum's Blockchain} 

{\color{black}

Traditionally, principals have controlled the oversight tasks of their agents, which now can be delegated to decentralized computer networks, with the following advantages:

\begin{itemize}
\item Blockchain technology provides a formal guarantee for principals and agents involved in solving agency problems in corporate governance. 
\item Blockchain technology can facilitate the elimination of agents as intermediaries in corporate governance through code, peer-to-peer connectivity, groups, and collaboration. 
\item DAO token holders are not affected by the existing corporate hierarchy and its restrictive effect. DAO token holder focuses on adding value, which benefits all components. 
\item The ``work value focus of workflow" in DAO structure has the potential to reform the agency relationship. 
\end{itemize}

DAO technology greatly contributes to improving the efficiency of agency relationships and reducing agency costs by an order of magnitude.

The purpose of Leonhard \textit{et al.} \cite{leonhard2017corporate} is to develop a virtual corporation with a corporate structure similar to the structure of a modern corporation. It offers a corporate governance structure where shareholders appoint the members of a board of directors, which then funds a Chief Executive Officer, who can then in turn pay salaries and acquire property on the corporation's behalf.

}

\subsubsection{Blockchain Governance—A New Way of Organizing Collaborations?} 

{\color{black}

Today, many companies are investing resources to develop and implement blockchain-based programs.

In the traditional contract governance, the effectiveness of contract management cooperation depends on the quality of national legal system to a large extent.

Conversely, DAO governance does not directly depend on the enforceability of external legal systems. Enforcement in a blockchain is achieved through prescribed code and algorithms, such as smart contracts. Even more, in governance, direct connections between collaborators are not required in a blockchain.

Thus, blockchain may be considered the first form of governance that truly leverages digital technology's computational- and data-based capabilities, well beyond traditional forms of social governance.

DAO, like other governance mechanisms, does not govern all types of transactions equally well.

Lumineau \textit{et al.} \cite{lumineau2020blockchain} believe that DAO governance can reduce searching, monitoring, and enforcement costs but tends, but often means relatively high design costs.

}

\subsubsection{eGov-DAO: A better government using blockchain based decentralized autonomous organization} 
The e-government system has greatly improved the efficiency and transparency of the government's daily operations. However, most existing e-government services are provided centrally and rely heavily on individual control. Highly centralized IT infrastructures are more vulnerable to external attacks.  Moreover, internal malicious users can easily compromise data integrity. In addition, relying on individuals to monitor some workflows makes the system error-prone and leaves room for corruption. In fact, both government and business services have been hacked multiple times, from ransomware to denial-of-service attacks. 
To address these challenges, Diallo \textit{et al.} \cite{diallo2018egov} suggest using blockchain technology and Decentralized Autonomous Organization to improve e-government systems. Diallo describes the high-level architecture of the government DAO and gives the detailed design of a DAO e-government system. This is the first system to allow real-time monitoring and analysis of e-government services.  The system retained all audit records, thereby limiting litigation between parties, increasing the speed of contract allocation and enforcement, providing transparency, accountability, immutability and better management of the national resources of the service. The evaluation of this system indicates that the government DAO system faces two main threats: data integrity and rule integrity. Besides, it can be found that modern government work does not require much delay. 
We can get conclusions that by implementing transparent and secure e-government systems at the lowest cost, eGov DAO's solution can help governments save unlimited resources, manage government businesses more effectively, and reduce the risk of providing contracts to companies that lack the capacity to deliver.

\subsubsection{Blockchain government: a next form of infrastructure for the twenty-first century} 

Today, there are hundreds of blockchain projects around the world to transform government systems.  There are signs that blockchain is a technology directly related to social organization. However, according to Jun \textit{et al.} \cite{jun2018blockchain}, there may be an epistemological rejection of the idea of blockchain-based automated systems replacing familiar public domains such as bureaucracy. 
Society must accept that such a shift is inevitable, and open discussion is needed to reduce the fear and side effects of introducing revolutionary new technologies. By applying Lawrence Lassig's ``code is law" proposition, the authors of \cite{jun2018blockchain} suggested that five principles should be followed when replacing bureaucracy with blockchain system: a) Introducing blockchain regulations; b) Transparent disclosure of data and source code; c) Implement independent executive management; d) Establish a governance system based on direct democracy; e) Make distributed autonomous government. 
Jun \textit{et al.}' proposed blockchain feature, which creates inviolable ``absolute laws", makes it possible to implement social technologies that can replace existing social institutions, including bureaucracy.

\subsubsection{The Potential for Blockchain Technology in Corporate Governance} 

{\color{black}

Traditional platforms are rapidly becoming obsolete and the trend is towards open platforms for financial services. Tech companies are starting to offer simple but disruptive financial services. In response, big financial firms are partnering with tech companies to maintain their market power.

Akgiray \textit{et al.} \cite{akgiray2019potential} discuss the latest applications of blockchain technology in financial services and outlined regulatory responses. Economists consider money to have three basic functions: a medium of exchange, a unit of account, and a store of value. It is easy to imagine a variety of DAO tokens in the financial world, because transactions are expressed in real currencies or digital currencies (Alipay, Apple Pay, etc.), and expressed in real money.

Therefore, Akgiray \textit{et al.} \cite{akgiray2019potential} argue, if the tokens on the DAO can fulfill the three functions of money, then the question of whether it is money or not can be raised.
}

\subsubsection{LikeStarter: a Smart-contract based Social DAO for Crowdfunding} 

Social media platforms are recognized as important media for the global transmission and dissemination of information. The combination of social interaction and crowdfunding represents a powerful symbiotic relationship. On the other hand, blockchain technology has revolutionized the way we think about the Internet. So Zichichi \textit{et al.} \cite{zichichi2019likestarter} introduced LikeStarter, a social network where users can raise money for other users through simple ``like" on the Ethereum blockchain. 
LikeStarter is built on the Ethereum blockchain, structured as a DAO that promotes crowdfunding without any central agency intervention and uses smart contracts to control and manage money. Zichichi \textit{et al.} \cite{zichichi2019likestarter} have used a case-study to show that LikeStarter successfully makes it easy for people to get funding and reach as many people as possible.

\subsubsection{Toward blockchain-based accounting and assurance} 

Since 2009, blockchain has become a potentially transformative information technology that promises to be as revolutionary as the Internet. Accounting and insurance could be one of the industries where blockchain could bring huge benefits and fundamentally change the current model. However, the potential benefits and challenges that blockchain could bring to accounting and assurance remain to be explored. 

For this reason, Dai \textit{et al.} \cite{dai2017toward} proposed an accounting and assurance method (an accounting ecosystem that supports blockchain, real-time, verifiable and transparent) based on the reference of multiple disciplines and ideas of the accounting industry. This method will provide real-time, verifiable information disclosure and step by step automation.
As a result, blockchain can be used as a tool to verify any information related to auditing. However, it is worth pondering how to adapt the existing blockchain mechanism to the field of accounting and auditing. The insights of Dai \textit{et al.} \cite{dai2017toward} will help integrate blockchain into the existing business processes and facilitate the transformation of the current audit model to the next generation. 

\subsubsection{Thinking outside the block: Projected phases of blockchain integration in the accounting industry} 

The rapid growth of blockchain has sparked curiosity across the industry, leading to talks about setting up a blockchain consortium in accounting. 
Accountancy firms such as PwC, Deloitte, EY and KPMG have pledged to integrate blockchain into their financial services. Innovative products such as Vulcan (for managing digital assets), Rubix (for improving supply change management), editable blockchain and blockchain as a service are among there.
Karajovic \textit{et al.} \cite{karajovic2019thinking} conducted an in-depth and detailed analysis on the application of blockchain technology in the accounting profession. As block linking becomes more mainstream, the technique can be used to simplify many redundant and vulnerable accounting practices. While the initial cost of developing and integrating blockchain infrastructure can be high, this can be offset by the cost savings it brings to the enterprise in terms of long-term, improved efficiency.
Karajovic\textit{et al.}'s philosophical analysis of the technology's application illustrates many questions about the uncertain relationship between accounting and blockchain. While the technology has the potential to reshape entire capital markets, the social and political barriers to blockchain proliferation need to be looked at critically. But one thing is certain: accounting is just one block in the chain that is being dramatically redefined by this disruptive technology.

\subsubsection{Blockchain based billing system for electric vehicle and charging station} 

{\color{black}

DAO can also be used in the charging system. The charging results measured in the charging electric vehicles may differ from the amount claimed in the charging station. This is because electric vehicles and charging station measure the charge amount with their own measurement equipment. If the electric vehicles or charging station provide fault information, the billing may be invalid. In addition, billing information can be manipulated. To prevent these problems, Jeong \textit{et al.} \cite{jeong2018blockchain} proposed a blockchain-based billing system. After mutual authentication, electric cars and charging stations both store billing information in the blockchain. A DAO is a system where all nodes have the same ledger, thus the ledger data cannot be tampered. Jeong \textit{et al.} have shown that the system prevents users from modifying their records after electric vehicles have been charged.

}

\subsubsection{A privacy-preserving voting protocol on blockchains} 

Voting is a universal phenomenon and to some extent is part of various societies. As the technology matures and more voting is expected in the future, there is an urgent need for a local voting mechanism built directly on the blockchain network to decentralize and dis-intermediate the network. Zhang \textit{et al.} \cite{zhang2018privacy} argue that the need for this voting mechanism is not limited to public blockchain networks, but also applies to syndicate-licensed blockchain networks.

In this regard, the authors categorized and summarized existing voting systems, and proposed a local voting mechanism based on blockchain to facilitate the decision-making of nodes in the blockchain network. The core idea is (a) distribution voting, (b) distributed tally, and (c) cryptography-based verification.
The implementation on the Hyperledger structure shows that the protocol is feasible for small and medium-sized voting issues. The agreement protects the privacy of voters and allows for the detection and correction of cheating without any credible parties.


\section{Open Issues}\label{sec:openissue}

{\color{black}

According to Diallo \textit{et al.} \cite{diallo2018egov}, we can see that although The DAO project was abolished, the future of Ethereum is bright. 
In the future, research on DAO can start from the following aspects. To help gain a quick clue, we propose several questions here.

Based on the system design of the blockchain itself, the first question is how to optimize and handle the performance trilemma, i.e., decentralization, security and scalability, in a balanced way in the context of DAO?
Secondly, a better blockchain system and the evaluation tool of blockchain security-enhancement solutions are anticipated for DAO. Furthermore, the feasibility of those blockchain system and evaluation tool needs to be assessed, because the governance is possible only if the benefits of efficiency, security and privacy of DAO exceed the social costs. Thus, another questions is how to reduce the social costs such that the feasibility of DAO can be improved.
}

{\color{black}

From the operation perspective of DAO, we have the following concerns.
\begin{itemize}
	\item Only depending on the social-optimal hard-fork rule is not applicable. The corporate governance solution based on DAO needs to propose new protocols for blockchain governance. 

	\item Leveraging DAO, it is necessary to create the trust of real transactions, thus maintaining the integrity of its contractual relationship.

	\item While ``code rules" can be formally followed on a particular chain. In exceptional states, sovereignty is asserted through an off-chain mechanism. This requires the study of possible actions, which are taken by the blockchain community to address anomalies in the ways that are consistent with their respective ideologies. 
 
	\item Legal issues related to DAOs need to be taken into account, such as how an entity should comply with the tax system. 
 
	\item Researchers need to explore the ultimate impact of DAO on human beings. The most powerful blockchain applications are likely to exacerbate inequality. Both researchers and engineers shall think about how to avoid and solve this issue.
\end{itemize}
	
 All of the mentioned concerns require a strong emphasis on interaction and communication between institutions and informal coding communities, thus it can help further research and develop new integration of social and organizational structures.
 }

{\color{black}

Currently, people are still maintaining a high expectation on DAO.
Governments around the world are betting that it will change the way they govern \cite{8368494}. 
For example, Dubai \cite{Dubai} sets a goal to ensure that by 2020 all government documents are stored in the blockchains. Other governments are studying its potential applicability in central banking, electronic voting, identity management and registry management. Blockchain transforms government operations to inspire new service delivery models for governments \cite{alketbi2020novel}.

DAO is believed to have many advantages over existing solutions.
With the emergence of Ethereum Classic \cite{8345547} and the incredible pace of new development, the platform is becoming matured \cite{jones2019blockchain}.
 In the infancy stage of DAO, smart contracts are bound to cause vulnerabilities like \textit{The DAO} attack, which will lead to better code checking mechanisms and secure coding practices to avoid such pitfalls. 
 In the future, not only may it be possible to establish full-fledged DAOs in multiple fields, it is also very likely to establish a unified single currency platform leveraging the technologies of DAO.
 }

\section{Conclusion}

{\color{black}
DAO is viewed as a very promising paradigm for our future decentralized organizational solutions.
This article reviews the most recent research activities on both academic and engineering scenarios, which basically include the governance problems and solutions, and typical DAO technologies and the related areas.
We hope that this overview can help researchers and engineers to identify the state-of-the-art studies of blockchain-based DAO.
}





\bibliographystyle{IEEEtran}
\bibliography{reference}

\begin{IEEEbiography}
[{\includegraphics[width=1in,height=1.25in,clip,keepaspectratio]{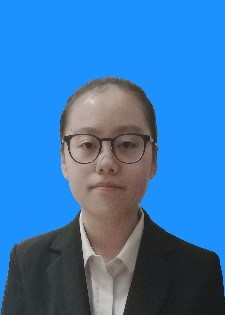}}]
{Lu~Liu} is currently an undergraduate with the School of Computer Science and Engineering, Sun Yat-Sen University, China. Her research interests include blockchain and financial big data.
\end{IEEEbiography}

\begin{IEEEbiography}
[{\includegraphics[width=1in,height=1.25in,clip,keepaspectratio]{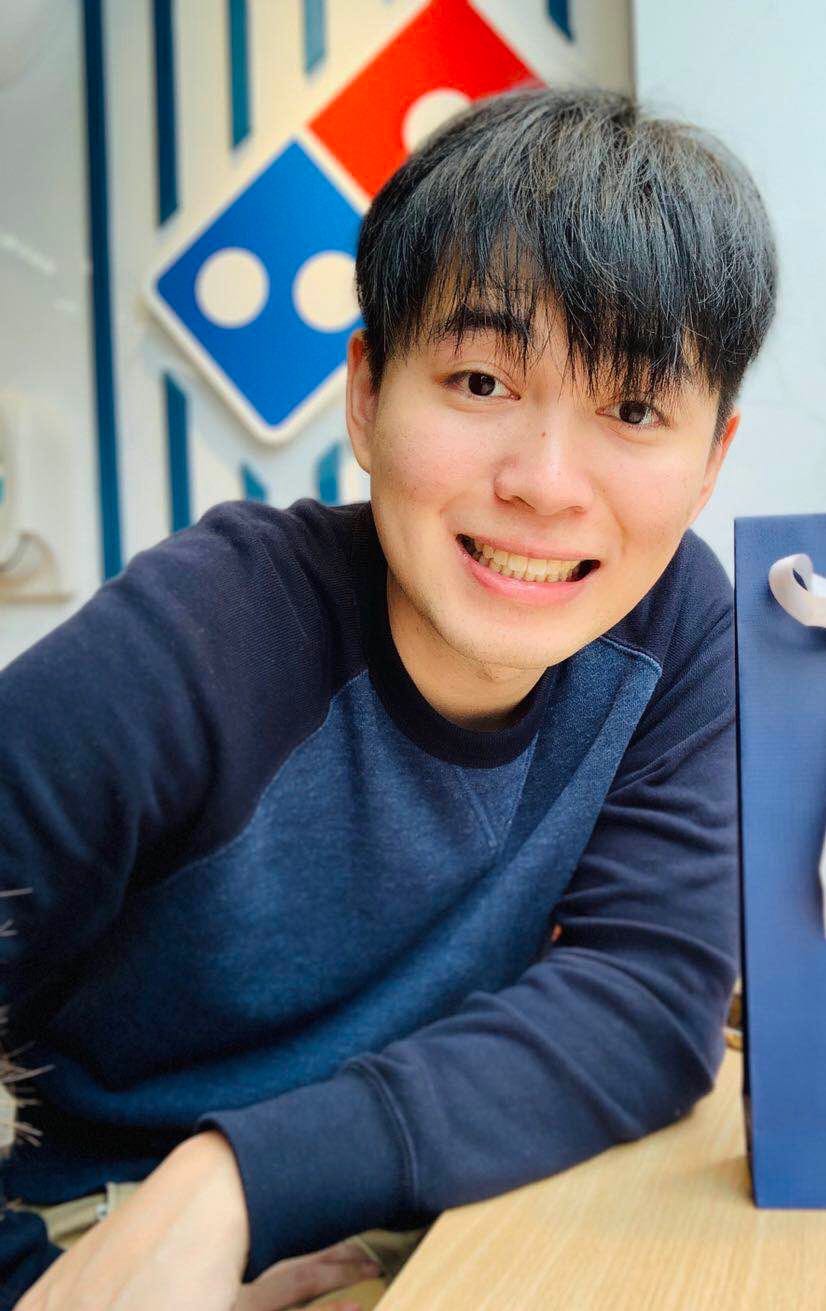}}]
{Sicong~Zhou} is currently a master-program student with the School of Computer Science and Engineering, Sun Yat-Sen University, China. His research interests mainly include distributed learning and blockchain performance optimization.
\end{IEEEbiography}

\begin{IEEEbiography}
[{\includegraphics[width=1in,height=1.25in,clip,keepaspectratio]{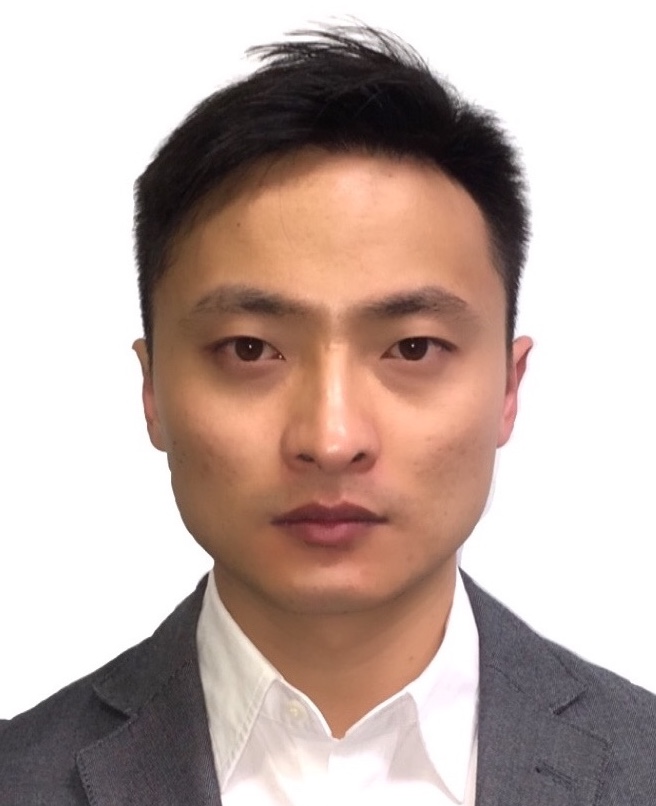}}]
{Huawei~Huang} (M'16) is currently an Associate Professor with Sun Yat-Sen University, China. He earned his Ph.D. degree in Computer Science and Engineering from the University of Aizu (Japan) in 2016. His research interests include blockchain and distributed computing. He has served as a Research Fellow of JSPS (2016-2018); a visiting scholar with Hong Kong Polytechnic University (2017-2018); an Assistant Professor with Kyoto University, Japan (2018-2019).  He received the best paper award from TrustCom2016. He is a member of ACM.
\end{IEEEbiography}

\begin{IEEEbiography}[{\includegraphics[width=1in,height=1.25in,clip,keepaspectratio]{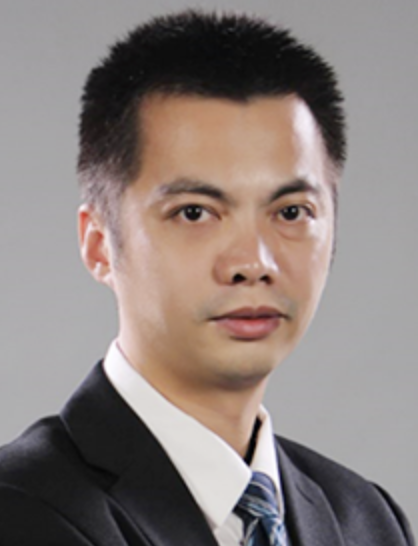}}]{ZIBIN~ZHENG} received the Ph.D. degree from the Chinese University of Hong Kong, in 2011.
He is currently a Professor at School of Computer Science and Engineering with Sun Yat-sen University, China. He serves as Chairman of the Software Engineering Department. He published over 120 international journal and conference papers, including 3 ESI highly cited papers. According to Google Scholar, his papers have more than 9600 citations, with an H-index of 46. His research interests include blockchain, services computing, software engineering, and financial big data. He was a recipient of several awards, including the Top 50 Influential Papers in Blockchain of 2018, the ACM
SIGSOFT Distinguished Paper Award at ICSE2010, the Best Student Paper Award at ICWS2010. He served as BlockSys'19 and CollaborateCom'16 General Co-Chair, SC2'19, ICIOT'18 and IoV'14 PC Co-Chair.

\end{IEEEbiography}

\end{document}